\documentclass{appolb}
\usepackage{epsfig}

\begin{document}
\title{ ON  TERAELECTRONVOLT MAJORANA NEUTRINOS}
\author{JANUSZ GLUZA
\address{
Department of Field Theory and Particle Physics, \\
Institute  of Physics, University of
Silesia \\
Uniwersytecka 4, PL-40-007 Katowice, Poland}
}

\maketitle

\begin{abstract}
The issue of existence of Majorana neutrinos with  masses of the order of TeV and  substantial couplings  is  addressed. 
A general neutrino mass matrix $M_{\nu}$ with both features is constructed, however, the form of  $M_{\nu}$ is constrained very much
by severe relations among the elements of $m_D$ and $M_R$ submatrices of  $M_{\nu}$. 
These general relations follow from the perturbative  construction of the light neutrino mass spectrum.  To avoid  such large correlations between low mass parameters in $m_D$ 
and large mass parameters in $M_R$, the Je\.zabek-Sumino see-saw model of bi-maximal neutrino mixing adopted to the TeV scale and the issue of possible symmetries of the matrix $M_{\nu}$ 
are  discussed. 
Results are supported by a few  numerical examples which show directly the complexity of the problem.
\end{abstract}

\section{Introduction}

In the see-saw scenario \cite{seesaw} light neutrinos with masses of the eV scale
demand heavy neutrino masses to be at least of the order of $10^{9}$ GeV. 
The diagonalization of  the neutrino mass matrix

\begin{equation}
M_{\nu} =  \left( 
\matrix{
0 & m_D \cr
m_D^T & M_R }
\right) ,
\label{mnu}
\end{equation}

in the case  $m_D << M_R$ (we assume, without lose of generality, that both matrices 
are of dimension 3),  gives 3 light $m_\nu$ and 3 heavy $m_N$ masses of neutrinos.
To a good approximation their scale  is given by \cite{grimus} 

\begin{eqnarray}
m_\nu & \simeq & -m_D^T M_R^{-1} m_D, \label{light} \\
m_N & \simeq & M_R. \label{heavy}
\end{eqnarray}

We see that to get light neutrino masses of the order of electronvolts one needs elements of the matrix  $M_R$  larger 
than  $10^{9}$ GeV (elements of the $m_D$ matrix are typically taken to be of 
the order of 1 GeV, scale of masses of charged leptons). Smaller masses of light neutrinos demand even larger $M_R$. 

It is clear that heavy neutrino states  exhibit huge masses. Not only this, but their couplings to the ordinary matter are negligible, namely,
if $m_D << M_R$,  the light-heavy neutrino mixing matrix $U^{Lh}$, which is a part of
the full unitary matrix U diagonalizing 
$M_{\nu}$ $ \left( U^T M_{\nu} U = m_{diag} \equiv diag \left[ m_i ,  M_i \right] \right) $ defined as

\begin{equation}
U=\left( 
\matrix{
{(U^{Ll})}^{\ast} & {(U^{Lh})}^{\ast} \cr 
U^{Rl} &  U^{Rh} }
\right) 
\label{mieszdef}
\end{equation}
exhibits very small elements

\begin{equation}
U^{Lh} \sim m_D M_R^{-1} \ll 1.
\label{miesz}
\end{equation}

In Eq.~\ref{mieszdef} the $U^{Ll}$ submatrix is responsible for the neutrino mixing in the light sector, 
while the submatrix $U^{Rh}$ describes neutrino mixings in the heavy neutrino sector. More details can be found  for instance in \cite{not,1993}.
If we look now into the form of the SM purely left-handed charged current written out in the mass eigenstates basis
\cite{1993} ($\nu_i$ and $N_i$ corresponds to light (heavy) neutrino mass states $m_i$ ($M_i$), respectively)

\begin{eqnarray}
\mathcal{L}_{CC}&=&
\frac g{\sqrt{2}} \left[ \sum\limits_{i=1}^{3} \overline{\mathcal{\nu }}_i \left( U^{Ll} \right)_{il} \gamma ^{\mu} P_L {l}W_\mu^{+} +
 \sum\limits_{i=1}^{3}   \overline{N}_i \left( U^{Lh} \right)_{il}  \gamma ^{\mu  %
} P_L {l}W_{\mu  }^{+} \right]  + h.c., \nonumber \\
&&  \label{cc} 
\end{eqnarray}
it is obvious that effects from the heavy neutrino sector on processes with charged currents are completely unimportant
(the same is true for the neutral   
current interactions \cite{1993}). This is a typical situation when the see-saw mechanism is
explored.
However, from experimental data we only know, that neutral leptons with masses below around ${\cal{O}}(10^2)$ GeV and with the typical weak neutrino coupling 
strength $g$ are excluded \cite{exphn}. There is no direct information on heavier neutral particles.
From  global fits to the data some bounds on the mixings of heavy neutrinos have been obtained \cite{glob}
  
\begin{eqnarray}
\sum\limits_{N} \mid (U^{Lh})_{Ne} \mid^2 & \leq & 0.0054 , \label{kappa_e} \\
\sum\limits_{N} \mid (U^{Lh})_{N \mu (\tau)} \mid^2 & \leq & 0.0028 (0.016).
\label{kappa_mutau}
\end{eqnarray}

These numbers are not negligible and effects of heavy neutrinos physics with the above mixings could be detected 
in future lepton  (e.g. $e^+e^-$ \cite{epem}, $e^-e^-$ \cite{emem,duka,bel}) or hadron \cite{had} colliders.
They can also influence processes generated by higher order corrections \cite{loop}. 
Finally, they may modify neutrino oscillation phenomena \cite{osc}.

However, a natural question arises:
is there any natural mechanism of heavy neutrinos creation with reasonably large mixings, or more precisely, what is the form of 
$M_{\nu}$ which would give such neutrino properties?
 Obviously, heavy neutrinos with TeV masses may lead to large $U^{Lh}$ elements, but too large masses of light neutrinos $m_i$ 
 would simultaneously arise.
Usually symmetry arguments are invoked to show that it is possible to build up an appropriate form of $M_{\nu}$ \cite{symm}. 
There are also other scenarios which implement TeV neutrinos. They go  
in different ways, e.g.: charged Higgs bosons \cite{chb}, bulk neutrinos or scalars
\cite{bn}, higher dimensional operators \cite{hdo}, naturally suppressed Dirac masses (through the presence of an extra scalar doublet) 
\cite{ma}. Whether any of these scenarios  can be really  assumed to be ``natural'' (and, maybe, used by 
nature) is an open question.

Here we would like to present and discuss the issue paying special attention to numerical results and their 
consequences. In the next section three typical examples of possible $M_{\nu}$ 
are given. In the first case masses of light neutrinos are too large, in the second  $M_{\nu}$
is constructed to give exactly 3 massless neutrinos and finally appropriate masses of light neutrinos
are obtained.
At the same time masses of heavy neutrinos in the range  ($100 \; GeV  \leq M_i \leq 1$ TeV)
and large $U^{Lh}$ elements which fulfill basic bounds 
Eqs.~\ref{kappa_e},\ref{kappa_mutau} are obtained.
Two lepton flavour violating  (LFV)  processes, namely neutrinoless double beta
and $\mu \to e \gamma$ decays  are considered and the issue of 
large neutrino mixings in the light sector is discussed.
Section 3 includes a discussion of a model with independent $m_D$ and $M_R$ matrices and the issue of possible symmetries of the matrix $M_\nu$ which could lead to the 
desired features of heavy neutrinos.

In the numerics we have to deal with huge differences of scales ($M_i/m_i \geq 10^{11}$) and the precision of calculations must be under control. 
To tackle this problem we used  MATHEMATICA \cite{math}.
Simple cross checks of calculations are: unitarity of the U  matrix Eq.~\ref{mieszdef} 
and recovering of  $M_{\nu}$  
to the same order of precision by the reverse relation 
$\left[ U m_{diag} U^T = M_{\nu} \right]$. 
All  results are obtained for the case of the $6 \times 6$ matrix $M_{\nu}$.

\section{TeV neutrino mass models with correlations among $m_D$ and $M_R$ matrix elements}

\underline{Example I: too large masses of light neutrinos}

Let us start from the most general case where we only assume  the naturalness of the $m_D$ and $M_R$ scales, without 
deeper insight into the relation
among their elements. Let us take then the elements of the $M_{\nu}$ mass matrix Eq.~\ref{mnu} in the following form

\begin{eqnarray}
m_D&=&\left( 
\begin{array}{ccc}
0.8 & 1 & 0.9 \\ 
1.5 & 0.5 & 0.1 \\ 
0.7 & 1.2 & 2
\end{array}
\right) \mbox{\rm [GeV]},  \label{mdir1} \\
&& \nonumber \\
M_R&=&Diag(100,150,200)\;  \mbox{\rm [GeV]}  \label{mr1}.
\end{eqnarray}

Without loss of generality, $M_R$ has been taken in a diagonal form 
\cite{diag} with its elements close 
to the present experimental limit \cite{exphn}. 
The result is:

\begin{eqnarray}
m_{diag}& \simeq &Diag(3.5 \cdot 10^{-4}, 0.013, 0.062 ,100,150,200)\; \mbox{\rm [GeV]}, \label{mdiag1} \\
&& \nonumber \\
U^{Ll} \ & \simeq & i \cdot
\left( \matrix{ 
0.854   & -0.022  & 0.519   \cr
-0.28   & 0.822   & 0.496   \cr
-0.438  & -0.57    & 0.695  \cr
} \right),\;\;\; \nonumber \\
U^{Lh} & \simeq &  \left( \matrix{ 0.008 & -0.007 & -0.005 \cr
0.015 & 0.003 & -0.001 \cr
0.007 & 0.007 & -0.01} \right).
  \label{ag1} 
\end{eqnarray}

In the above we restrict ourselves to the only interesting cases of 
light-light ($U^{Ll}$) and light-heavy ($U^{Lh}$) neutrino mixing sectors.
We can see that large  LH neutrino mixings can be obtained
but the masses of light neutrinos are too large. 
It is clear now, after 
neutrino oscillation data analysis that the mass of the heaviest of light neutrinos must be in the range \cite{bar}
\begin{equation}
0.04\mbox{\rm  eV}\leq m_{3 }\leq 2.7\mbox{\rm  eV}.
\label{m3}
\end{equation}
Let us note that both mixing angles and the mass spectrum still satisfy the see-saw  relations  Eq.~\ref{light}  and Eq.~\ref{miesz}.

Similarly it can be checked that the LH mixings could be larger with  larger $m_D$ elements.  However, larger  $m_D$ would increase masses of light neutrinos 
(in agreement with Eq.~\ref{light}).
Taking smaller   $m_D$ would on the other hand lead to  appropriate $m_i$, but
this time LH mixings would be completely out of interest. So, we need $m_D$ elements to be at 
${\cal{O}}(1)$ GeV level.

We can still try to change the ranks of $m_D$ and $M_R$ matrices. Then we can expect that additional light  neutrinos appear.
First, let us  put the first  entry in Eq.~\ref{mr1} as zero ($100 \to 0$) without other changes in the $m_\nu$ matrix elements.
The result is

\begin{equation}
m_{diag} \simeq Diag(3.8 \cdot 10^{-4}, 0.0228, 1.83 , 1.85 ,150,200)\; \mbox{\rm [GeV]}. \label{mdiag1aa} 
\end{equation}

Second, let us also take the second entry  in Eq.~\ref{mr1} as zero ($150 \to 0$). Then the mass spectrum is

\begin{equation}
m_{diag} \simeq Diag( 0.001, 0.783, 0.792 , 2.33 , 2.34 ,200)\; \mbox{\rm [GeV]}. \label{mdiag1aaa} 
\end{equation}
 
We can see  that in the first case we get neutrinos with two masses of the order $m_D/M_R$, two of the order $m_D$ and two of the order 
$M_R$. In the second case these are: one mass of the order $m_D/M_R$, four masses  of the order $m_D$ and one of the order 
$M_R$. This result fits to the discussion of the dependence between the rank of the $M_R$ matrix and the scale of the obtained neutrino masses as given 
in \cite{lin}. 

We can also decrease the rank of the matrix $m_D$. For that we take as an example 

\begin{eqnarray}
m_D&=&\left( 
\begin{array}{ccc}
0 & 0 & 0 \\ 
0 & 0 & 0 \\ 
0.8 & 0.9 & 1
\end{array}
\right) \mbox{\rm [GeV]},  \label{mdir1a} \\
&& \nonumber \\
M_R&=&Diag(0,150,200)\;  \mbox{\rm [GeV]}  \label{mr1a}.
\end{eqnarray}

The $m_D$ matrix is of rank 1, the   $M_R$ matrix has  the rank 2. 
The result is:

\begin{eqnarray}
m_{diag}& \simeq &Diag(0, 0, 0.793 , 0.807 ,150,200)\; \mbox{\rm [GeV]}, \label{mdiag1a} \\
&& \nonumber \\
U^{Ll} \ & \simeq & 
\left( \matrix{ 
1  & 0 & 0\cr
0 & 1  & 0  \cr
 0 & 0  & 0.704 \cr
} \right),\;\;\; 
U^{Lh} \simeq  \left( \matrix{ 0 & 0 & 0 \cr
0 & 0 & 0 \cr
-0.71 & -0.009 & 0.005} \right).
  \label{ag1a} 
\end{eqnarray}

Two massless neutrinos are obtained.
Though the rank of the matrix $m_D$ is 1, we get two neutrinos with masses of the order of $m_D$ elements. 
We can not further decrease the rank of the $m_D$ matrix as  heavy neutrinos would not mix with light states at all. 
Taking only one nonzero entry in the $M_R$ matrix would not change the situation. The neutrino mass spectrum includes in this case
three massless neutrinos, two massive neutrinos of the order of $m_D$ and one heavy neutrino. 

In this way we have shown that changing ranks of $m_D$ and $M_R$ matrices is not sufficient to get appropriate spectrum of neutrino masses
(for further discussion of the meaning of $m_D$ and $M_R$ matrices of different ranks in  general see-saw models see e.g.  \cite{rank}).

We can see that there is no way around and we have to look for some relations among $m_D$ and $M_R$ elements of the 
$M_{\nu}$ matrix which would
give appropriate masses of light neutrinos and  TeV neutrinos with substantial mixings. 

\underline{Example II: 3 massless neutrinos}

Let us take \cite{had}

\begin{equation}
m_D=m_D^{(0)}+m_D^{(1)},
\label{mdirogolna}
\end{equation}
and assume that $m_D^{(0)} >> m_D^{(1)}$. 

Using       Eq.~\ref{light} we get

\begin{eqnarray}
m_{\nu  } &=&
-m_D^{(0)}\frac 1{M_R}m_D^{(0)T}-\left( m_D^{(1)}\frac
1{M_R}m_D^{(0)T}+m_D^{(0)}\frac 1{M_R}m_D^{(1)T}\right) \nonumber \\
& -& m_D^{(1)}\frac 1{M_R}m_D^{(1)T}\mbox{\rm  .}  
\end{eqnarray}

The first term is the largest. It will give  the largest contribution to $m_{light}$.
Let us demand that it is zero, i.e. $m_D^{(0)}\frac 1{M_R}m_D^{(0)T}=0$ and parameterize $m_D^{(0)}$
in the most general way (elements of Eq.~\ref{mdiraczero} can be complex)

\begin{equation}
m_D^{(0)}=\left( 
\begin{array}{lll}
\alpha_1 & \alpha_2 & \alpha_3 \\ 
\beta_1 &  \beta_2 & \beta_3 \\ 
\gamma_1 & \gamma_2 & \gamma_3
\end{array}
\right) \equiv  
\left( \matrix{ \alpha_i \cr
                \beta_i \cr
                \gamma_i } \right), \;\;\;i=1,2,3.
\label{mdiraczero}
\end{equation}



Then the following relation can be obtained ($M_R$ has diagonal elements $M_1,M_2,M_3$)

\begin{equation}
\sum\limits_i \left( \matrix{ \frac{\alpha_i^2}{M_1} & \frac{\alpha_i \beta_i}{M_2} & \frac{\alpha_i \beta_i}{M_3} \cr
                \frac{\alpha_i \beta_i}{M_1} & \frac{\beta_i^2}{M_2} & \frac{ \beta_i \gamma_i}{M_3} \cr
                \frac{\alpha_i \gamma_i}{M_1} & \frac{\beta_i \gamma_i}{M_2} & \frac{ \gamma_i^2}{M_3} }
       \right) =0.
\label{soph}
\end{equation}
If $m_D^{(1)}=0$, it is a set of equations for relations among $m_D^{(0)}$ and $M_R$ elements which 
assure that three massless neutrinos are constructed. 
To show a numerical example, we will leave for a  moment the most general case Eq.~\ref{mdiraczero}
and use $m_D^{(0)}$ with  the following texture 
 
\begin{equation}
m_D^{(0)}=\left( 
\begin{array}{lll}
\alpha_i  \\ 
a \cdot \alpha_i \\ 
b \cdot \alpha_i
\end{array}
\right) \mbox{\rm  .}  
\label{simpler}
\end{equation}

Then, instead of Eq.~\ref{soph} we get only one condition

\begin{equation}
\frac{\alpha_1^2}{M_1}+\frac{\alpha_2^2}{M_2}+\frac{\alpha_3^2}{M_3}=0\mbox{\rm  .}
\label{kappawarunek}
\end{equation}

Now we will use also 
the second important relation among  heavy neutrino mass matrix elements  which comes from 
the neutrinoless double beta experiments \cite{bb}

\begin{equation}
\left| {\sum\limits_{i}  }(U^{Lh})_{ie}^2\frac 1{M_i}\right| =
{\omega }^2\mbox{\rm  ,}
\label{rozpadbeta}
\end{equation}
where ${\omega }^2<(2-2.8)\times 10^{-5}$ TeV$^{-1}$. There is no consensus concerning estimation of the
$\omega$ parameter, nevertheless it appear that this relation is so severe that the possibility  
of  heavy neutrinos detection in future colliders is drastically reduced \cite{bel} (see, however,
\cite{duka,mad}). It can be checked that conclusions of the present paper do not change when 
$\omega=0$ is assumed. Then relations Eq.~\ref{mr1},\ref{kappawarunek},\ref{rozpadbeta} with 
$\alpha_1=3$ GeV allow to set $\alpha_2$ and $\alpha_3$. $U^{Lh}$ is taken in the form of
Eq.~\ref{miesz}, $a=1$, $b=0$.
Then the matrix in Eq.~\ref{mnu}
is fixed and, after its diagonalization, the set of physical neutrino parameters is obtained

\begin{eqnarray}
m_{diag}& \simeq &Diag(0,0,0,100,151,201) \mbox{\rm [GeV]} \label{mdiag2} \\
&& \nonumber \\
U^{Ll} & \simeq & 
{
\left( \matrix{ 
0.998 & 0 & 0 \cr
-0.004  &  0.998 & 0 \cr
0  & 0 & 1. } \right) }, \;\;\;
U^{Lh}  \simeq  
{
\left( \matrix{ 
0.03 & -0.048 \cdot i & 0.036 \cr
0.03  & -0.048 \cdot i  & 0.036 \cr
0  & 0  & 0} 
\right)}.  \nonumber \\
&& \label{ag2} 
\end{eqnarray}

The LH neutrino mixing is large (Eq.~\ref{ag2}) and fulfill 
Eqs.~\ref{kappa_e},\ref{kappa_mutau}.
Three massless  neutrinos are there. The spectrum of heavy states is as expected.
In the SM massless neutrinos give diagonal $U^{Ll}$. Here some small nondiagonal entries reflect the
existence of heavy neutrino states.
Crucial is the Eq.~\ref{kappawarunek}.
If we disturb it slightly then some of light neutrino states exhibit unacceptable values,
e.g. if $\alpha_1 \to \alpha_1 + 10^{-6}$  GeV  then  (with the other parameters chosen  just as before) we get 

\begin{equation}
m_{diag}=Diag(0,0,2 \cdot 10^{-7},100,151,201) \mbox{\rm [GeV]}.
\end{equation}

\underline{Example III: realistic light neutrino masses}

To make our construction realistic two final issues  must be addressed. The first is the 
exact spectrum of light neutrino masses and the second is their mixing pattern.

As for the light neutrino parameters
let us try to recover bi-maximal mixings where \cite{exphn}

\begin{eqnarray}
  U_{\rm MNS} & \simeq & \left( 
    \matrix{ \frac{1}{\sqrt{2}} & \frac{1}{\sqrt{2}} & 0 \cr
     -\frac{1}{2} &  \frac{1}{2} & \frac{1}{\sqrt{2}} \cr
      \frac{1}{2} & -\frac{1}{2} & \frac{1}{\sqrt{2}} \cr } \right), \nonumber \\
&& \nonumber \\
 \Delta m^2_{\rm atm} & \simeq & (1.6 \div 4)\times 10^{-3}\mbox{\rm  eV}^2, \nonumber \\
&& \nonumber \\
 \Delta m^2_{\odot}  & \simeq & 10^{-2} \Delta m^2_{\rm atm}.
\label{bi}
\end{eqnarray}
The solar neutrino parameter $\Delta m^2_{\odot}$ is realized by the LMA-MSW scenario of neutrino oscillations. 

We take

\begin{equation}
m_D^{(1)}=U_{MNS}  \cdot  Diag \left( \matrix{ 0, & 10^{-11},&  8 \cdot 10^{-10}} \right). 
\end{equation}

Other parameters are the same as in Example II.
This sort of additional contribution to the Dirac mass $m_D$
does not  affect the heavy neutrino sector.
The result is 
  
\begin{eqnarray}
m_{diag}& \simeq &Diag(0,2 \cdot 10^{-11}, 6 \cdot 10^{-11} ,100,151,201) \mbox{\rm [GeV]}, \label{mdiag3} \\
&& \nonumber \\
U^{Ll} &\simeq & 
\left( \matrix{ 
0.577-0.01 \cdot i & -0.706-0.006 \cdot i & 0.003+0.405 \cdot i \cr
-0.577+0.01 \cdot i  & 0.001 & 0.007+0.814 \cdot i \cr
0.577+0.01 \cdot i  & 0.706+0.018 \cdot i & -0.01+0.409 \cdot i } \right), \;\;\; \nonumber \\
&& \nonumber \\
U^{Lh} & \simeq  & 
\left( \matrix{ 
0.03 & -0.048 \cdot i & 0.036 \cr
0.03  & -0.048 \cdot i  & 0.036 \cr
0  & 0  & 0} 
\right).  \label{ag3} 
\end{eqnarray}

The masses gives appropriate $ \Delta m^2_{\rm atm}$ and
 $\Delta m^2_{\odot}$ (Eq.~\ref{bi}). Large mixings in the $U^{Ll}$ sector are obtained. Of course, this matrix
is not unitary and differs from  $U_{MNS}$ in Eq.~\ref{bi}, the effect expected as  heavy neutrino states
affect the light sector. 

Finally let us comment on the  $\mu \to e \gamma $ decay. Analyses of experimental data give \cite{exphn}
 
\begin{eqnarray}
BR(\mu \rightarrow e\gamma )&=&
\frac{3\alpha }{8\pi }\left| 
\sum\limits_{i} (U^{Lh})_{ei}(U^{Lh})_{i\mu}^{\dagger} \frac{M_i^2}{M_W^2} 
 \right| ^2  \leq  4.9\times 10^{-11}.
\label{BRwzor}
\end{eqnarray}
In Eq.~\ref{BRwzor} contributions from the light neutrinos have been safely neglected \cite{loop}.
Taking into account  Eq.~\ref{ag3}, $BR(\mu \rightarrow e\gamma ) \simeq 1.4 \cdot 10^{-12}$.
It fits to the present limit.

\section{TeV neutrino mass models without fine-tuning problems and symmetry arguments}

Is it possible to avoid the problem of strong correlations among $m_D$ and $M_R$ elements (Eq.~\ref{soph} and Eq.~\ref{kappawarunek})?
As discussed  in \cite{jez}, $m_D$ and $M_R$ originate from apparently disconnected mechanisms of gauge symmetry breaking of the $SU(2) \times U(1)$
gauge group and some larger unification group, respectively. Thus, it is hard to believe that these are arranged  to fulfill Eq.~\ref{kappawarunek} 
just to  give TeV neutrinos with large LH mixings. In \cite{jez} a phenomenological model of the matrix $M_\nu$ with uncorrelated  $m_D$ and $M_R$ matrix 
elements which realizes bi-maximal neutrino mixing has been constructed. The result  discussed explicitly in \cite{jez} is the following

\begin{eqnarray}
m_D &=& m_3 \left( \matrix{ x^2 y & 0 & 0 \cr
0 & x & x \cr
0 & -x^2 & 1} \right), \label{mdj}  \\
M_R^{-1} (x=0) &=& \frac{1}{M} \left( \matrix{ 0 & 0 & \alpha \cr 0 & 1 & 0 \cr \alpha & 0 & 0} \right).
\label{mrj}
\end{eqnarray}

$m_3$ and $M$ are of the order of the top quark  and grand unification energy scale, respectively. $x={\cal{O}}(m_c/m_t)$ is of the order of $10^{-2}$ 
(the ratio of the charm and the  top quark mass), $y \simeq 10^{-1}$,
$\alpha < 1$. The relation $x=0$ in Eq.~\ref{mrj} stresses that $M_R$ is independent of $x$ being an element of $m_D$. Null matrix elements in  Eq.~\ref{mdj} and Eq.~\ref{mrj} are higher order
powers in $x$ and $y$ and are neglected. This model has been originally used in the context of see-saw  models. To accommodate it to TeV neutrinos,  the $M$ scale must be 
lowered to the TeV level. Then masses of light neutrinos of the order of $x^2 m_3^2/M$ appear (Eq.~31 in \cite{jez}) and, as in the previous section the fine-tuning problem shows up:
the numerator  must be tuned to fit light neutrino masses at the eV level. Moreover, 
using Eq.~\ref{miesz} we get

\begin{equation}
U^{Lh} \simeq \frac{m_3}{M} \left( \matrix{ 0 & 0 & 0 \cr x \alpha & x & 0 \cr \alpha & -x^2 & 0} \right).
\end{equation}

We can see that these mixings are negligible. There is no contribution to the neutrinoless double beta decay and to the $\mu \to e \gamma$ decay from heavy neutrino mixing.

Let us finally comment on   the possible symmetry of the full matrix  $M_{\nu}$. We  note  that the relations in Eq.~\ref{soph} and Eq.~\ref{kappawarunek} are not  symmetries of
$M_{\nu}$ but rather they are fine-tuning relations of  elements of $m_D$ and $M_R$ 
which ensures that appropriate masses of light neutrinos can be obtained
simultaneously with TeV neutrinos. 
Symmetries act directly on $M_{\nu}$ and not on objects which are
functions of elements of $M_{\nu}$. As discussed in the Introduction (and Example I), TeV neutrinos may lead to large LH mixings, but too large masses of light neutrinos 
 would simultaneously arise. A source of the problem lies in a different scale of the elements of  $M_{\nu}$. Could  symmetry of $M_\nu$  be able to reconcile the problem?
Let us consider a toy model with  only light $(\nu)$ and heavy $(N)$ neutrinos.
Let us assume that in the $\left( \nu, N \right)^T$ basis 
the neutrino mass matrix is (elements a,b,c are real numbers) 
\begin{equation}
M=\left( \matrix{ a & b \cr b& c } \right).
\end{equation}

The masses and a mixing angle are given by
\begin{equation}
m_{1,2}=\frac{1}{2}\left(a+c\mp\sqrt{(a-c)^2+4b^2} \right),
\end{equation}
and
\begin{equation}
\sin{2\xi}=\frac{2b}{\sqrt{(a-c)^2+4b^2}}.
\end{equation}

If  $c>>b,a$ then we get $|m_1| \simeq b^2/c$, $m_2 \simeq c \gg |m1|$ and $\xi \simeq b/c$. It is just a see-saw mechanism.
If, however,  $ac=b^2$ (due to symmetry!) then  $m_1 \simeq 0$, $m_2=a+c$ and $\sin 2 \xi = 2 \sqrt{ac}/(a+c)$. 
We can see that $\sin 2 \xi \simeq 1$ if $a \simeq c$, which is, however,  not a natural assumption.
The problem does not vanish with larger dimension of  $M_{\nu}$. To summarize,
a difficulty to build a symmetry of the $M_{\nu}$ matrix lies in the following: large LH mixings means that elements of the  $M_{\nu}$ matrix are 
comparable. However, this is not true as long as  the relation  $m_D<<M_R$ holds.

\section{Conclusions}

In summary, it has been shown that the present data from the light neutrino sector, especially their 
masses allow to construct  the neutrino mass matrix $M_{\nu}$ with both
TeV neutrinos and large LH mixings. However, the form of  $M_{\nu}$ is constrained very much. 
In all three numerical examples of section 2 LH mixings fulfill Eq.~\ref{miesz}. 
The kind of relations Eq.~\ref{soph} and Eq.~\ref{kappawarunek} do not change this fact.
There is no way around as long as $m_D<<M_R$ which is, as discussed in Example I, a condition which must be fullfiled 
for our purposes. 

The natural decomposition Eq.~\ref{mdirogolna} which has been used in this paper has been for the first time introduced in \cite{had}.  
At this time an information on neutrino masses was completely different. Muon and tau neutrino masses at the level of keV and MeV, respectively
has been allowed. Then some freedom of parameters in relations Eq.~\ref{soph} and Eq.~\ref{kappawarunek} was possible.

To avoid fine tuning problems we could look for a symmetry of the full matrix $M_\nu$  or build models with uncorrelated $m_D$ and $M_R$ matrices. However, as argued in section 3,
in the first case a kind of internal contradiction between a requirement of two different scales $m_D << M_R$ and large LH mixings arises. In the second case negligible 
LH mixings emerge. 

The basic  conclusions of the paper remain true regardless of the number of heavy neutrino states.
The case when left-handed fields give a Majorana mass term $M_L$ in Eq.~\ref{mnu}
\cite{duka,gun} come to the same class of basic fine-tuning problems.

\section*{Acknowledgments}
I would like to thank prof. M. Zra\l ek for useful discussions.
The work  was supported by the Polish Committee for Scientific Research under 
Grants No. 2P03B04919  and 2P03B05418.

\end{document}